\documentstyle{article}
\title{
\begin{flushright}
{\bf\normalsize   COLO-HEP-320}\\ \end{flushright}
\bf DGSOS on DTRS }

\author{ {\it C.F. Baillie} \\
	 Physics Dept. \\ University of Colorado\\ Boulder, CO 80309,
	 USA\\ \\ {\it W. Janke}\\ Institut fur Physik\\ Johannes
	 Gutenberg Universitat\\
	 Germany\\ \\ and
         \\ \\
         {\it D.A. Johnston}\\
         LPTHE\\
	 Universite Paris Sud, Batiment 211\\
         F-91405 Orsay, France$^{1}$\\
	 \\
         Dept. of Mathematics\\
         Heriot-Watt University\\
         Edinburgh, EH14 4AS, Scotland$^{2}$ }

\begin{document} \maketitle
%-----------------------------------------------------------------------
		      {\Large \begin{abstract}
%-----------------------------------------------------------------------
%
We perform simulations of a discrete gaussian solid on solid (DGSOS)
model on dynamical $\phi^3$ graphs, which is equivalent to coupling the
model to 2d quantum gravity, using the cluster algorithms recently
developed by Evertz et.al.for use on fixed lattices. We find evidence
from the growth of the width-squared in the rough phase of KT-like
behaviour, which is consistent with theoretical expectations.  We also
investigate the cluster statistics, dynamical critical exponent and
lattice properties, and compare these with the dual XY model.  \\ \\
Submitted to Phys Lett B.  \\ \\ \\ \\ \\ \\ \\ \\
$1$ {\it Address Sept. 1993 - 1994} \\
$2$ {\it Permanent Address}
%
%-----------------------------------------------------------------------
			\end{abstract} }
%-----------------------------------------------------------------------
%
  \thispagestyle{empty}
%
%***********************************************************************
%
  \newpage
%
%-----------------------------------------------------------------------
		  \pagenumbering{arabic}
%-----------------------------------------------------------------------

\section{Introduction} Following the work in \cite{1,2} in the
continuum Liouville theory formalism and the vast recent output of
results in the context of matrix models \cite{2a} it is now clear that
one can calculate the critical exponents for various $c \le 1$
conformal field theories coupled to 2d quantum gravity given the
conformal weights of operators in the ``bare'' theories without
gravity. The coupling to 2d quantum gravity can be incorporated in
discrete simulations of such models by having the matter live on, and
interact with, a dynamical lattice - typically a dynamical
triangulation or the dual dynamical $\phi^3$ graphs.  The Ising model
can be solved exactly on dynamical $\phi^3$ and $\phi^4$ graphs
\cite{3}, giving exponents in agreement with the continuum approach,
and these results are also backed up by numerical simulations, as are
the exponent values for 3 and 4 state Potts models coupled to 2d
gravity \cite{4}.  For $c>1$ the matrix models can no longer be exactly
solved and the continuum approach breaks down, but numerical
simulations show no obvious pathologies and neither do extrapolations
from exactly evaluated partition functions for small numbers of points
\cite{4a,4b}.

In view of this the borderline case of $c=1$, exemplified by the 4
state Potts model or the $XY$ model, is of particular interest. On a
fixed lattice the $XY$ model, whose partition function is
\begin{equation} Z= \prod_i \int d \theta_i \exp \left( \beta
\sum_{<ij>} \cos ( \theta_i -\theta_j) \right), \label{e01}
\end{equation} where the $\theta_i$'s are angular variables, is known
to undergo a Kosterlitz-Thouless (KT) transition driven by the
unbinding of vortex pair configurations \cite{5}.  The correlation
length diverges as \begin{equation} \xi = A_\xi \exp \left( { B_\xi
\over (T - T_c)^\nu } \right) \label{e02} \end{equation} and the spin
susceptibility as \begin{equation} \chi = A_\chi \exp \left( { B_\chi
\over (T - T_c)^\nu } \right), \label{e03} \end{equation} where the
exponent $\nu$ is predicted to be $1/2$.  The KT theory also predicts
the correlation function critical exponent $\eta = 1/4$, where $\eta$
is given by \begin{equation} \chi = C \xi^{2-\eta}.  \label{e04}
\end{equation} It has proved notoriously difficult to confirm this
behaviour numerically \cite{6} though there is now a general consensus
that the data supports a KT transition rather than possible
alternatives such as a second order transition. On a dynamical lattice
the theoretical predictions are that a KT transition persists \cite{7}
and the two simulations carried out to date appear to support this
assertion \cite{8,8a}.

Another class of lattice models, namely the solid on solid (SOS)
models, are expected to display a KT roughening transition. Different
variants of these models exist one of which, the body-centred solid on
solid (BCSOS) model, is equivalent to a six vertex model which is
exactly soluble and is known to have a KT transition \cite{9}.
Numerical evidence suggests that on fixed lattices other variants such
as the dual of the $XY$ model, the discrete gaussian solid on solid
(DGSOS) model we consider here and the absolute value solid on solid
(ASOS) model are in the same universality class \cite{10}. The
partition function for the DGSOS model on a fixed lattice is given by
\begin{equation} Z =  \sum_{h} \exp \left( - { \beta  } \sum_{<i,j>}
(h_i - h_j)^2 \right) \label{e05} \end{equation} where the $h$'s are
integer heights at each lattice point and the sum in the exponent is
over the edges in the lattice. In the ASOS model this is modified to
\begin{equation} \sum_{h} \exp \left( - \beta \sum_{<i,j>} |h_i - h_j|
\right), \end{equation} in the dual of the $XY$ model to
\begin{equation} \sum_{h} \prod_{<i,j>} I_{|h_i - h_j|}  ( \beta )
\end{equation} where $I_{|h_i - h_j|}$ is a modified Bessel function,
and in the BCSOS model to \begin{equation} \sum_{h} \exp \left( - \beta
\sum_{[i,j]} |h_i - h_j| \right) \end{equation} where $i,j$ are now
diagonal neighbours and nearest neighbours are constrained by $|h_i -
h_j|=1$.  The models have a rough phase at low $\beta$ separated from a
smooth phase at high $\beta$ by the $KT$ transition.  The width-squared
of the surface, which can be defined by \begin{equation} \sigma^2 = <
{1 \over N} \sum_i ( h_i - \bar h )^2 >, \label{e06} \end{equation}
where $\bar h$ is the mean value of the $h_i$, is predicted by KT
theory to diverge logarithmically in the rough phase with the number of
points $N$:  \begin{equation} \sigma^2 = {T_{eff} \over  d \pi } \log (
N ) + B, \label{e07} \end{equation} where $d$ is the fractal dimension
of the lattice (we have substituted $L = N^{1/d}$).  At the critical
point $\beta = \beta_c$ KT predicts $T_{eff} = 2 / \pi$ and as $\beta$
approaches $\beta_c$ from below $T_{eff} - 2 / \pi \simeq ( \beta_c -
\beta )^{1/2}$. On fixed lattices it appears that it is possible to
unambiguously verify equ.(\ref{e07}) for sufficiently small $\beta$
but that the behaviour of $T_{eff}$ as $\beta \rightarrow \beta_c$
cannot be fitted  without including corrections in the formula for the
width-squared coming from using a ``running temperature'' in the KT
flow equations \cite{10}.

\section{The Simulation Method}

In this paper we simulate the DGSOS model on dynamical $\phi^3$ graphs
of spherical topology with a fixed number $N$ of points and no tadpole
or self-energy insertions which would correspond to degenerate
triangulations on the direct lattice.  The partition function is now
\begin{equation} Z = \sum_{G^{(N)}} \sum_{h} \exp \left( - { \beta }
\sum_{<i,j>} G^{(N)}_{ij} (h_i - h_j)^2 \right) \label{e08}
\end{equation} where $G^{(N)}_{ij}$ is the connectivity matrix of the
graph.  Our aims are threefold:  \begin{itemize} \item{} We wish to see
if the KT predictions are still valid when the SOS model is coupled to
2d quantum gravity, as they appear to be in the $XY$ model case;
\item{} We wish to investigate the efficiency of the cluster algorithms
used in simulating the SOS models on dynamical lattices; \item{} We
wish to see how the lattice characteristics of the SOS model compare
with the Potts and $XY$ models simulated previously.  \end{itemize} To
this end we simulate graphs with
$N=100,200,300,400,500,1000,2000,5000,10000$ points for a range of
$\beta$ values from 0.05 to 5.0. For each data point we carried out
10,000 metropolis equilibration sweeps, followed by 50,000
measurements. Before each set of measurements we carry out a number of
Wolff updates using the $H$ and $I$ algorithms of Evertz et.al.  tuned
so that $N_{updates} = 1 /$ Cluster Size, as well as $N$ local ``flip''
moves in the lattice. Test runs and our experience with simulating the
Potts and $XY$ models on dynamical lattices provided assurance that
this was sufficient to allow for the lattice and the spin model to
interact.  To ensure detailed balance it is necessary to check that the
rings at either end of the link being flipped have no links in common.
The starting graphs came from the Tutte algorithm used to generate pure
two-dimensional gravity meshes.

The cluster algorithms used in the simulations are of the so-called
``valley to mountain reflection'' type \cite{10}. One chooses a
reflection plane at height $M$ and notes that all of the $h_i$ may be
written as $h_i = \sigma_i | h_i - M | + M$ where the $\sigma_i = \pm
1$ are embedded Ising variables determining whether the height $h_i$ is
above ($+$) or below ($-$) $M$.  A cluster is then built by choosing a
seed point and adding further links $<ij>$  with the probability
\begin{equation} P_{add} = 1 -  q \exp \left( - \beta | h_i - M | | h_j
- M | (\sigma_i \sigma_j + 1) \right) \label{e09a} \end{equation} where
$q \le 1$. It is then flipped by reversing the sign of the embedded
Ising variables in the cluster. The fixed lattice simulations in
\cite{10} revealed that the choice of the reflection plane $M$ was
crucial for the effectiveness of the algorithm, because the
width-squared of the surfaces was not that great (in the region of
$1-2$) for values of $\beta$ around $\beta_c$ even for large lattices.
One has to be careful to ensure that $M$ is not picked too far away
from the surface.  If we take $q=1$ in equ.(\ref{e09}) taking the
reflection plane to pass through the seed point will generate only
clusters of size 1 (``monomers'') so we must choose some other point
close to the seed. One possibility is to take  $M = h_{seed} \pm 1/2$,
where the plus and minus are chosen with equal probability, which was
called the $H$ algorithm ($H=$ half-integer) in \cite{10}. It was found
that this algorithm, while ergodic, still had a dynamical critical
exponent $z \simeq 1$.  It was pointed out in \cite{10} that single
step islands, configurations where a set of points on the surface were
a step above or below the background, could be created and destroyed by
the $H$ algorithm but not reflected by it. Such reflections would cost
nothing in terms of the Boltzmann weights and are thus likely to be
important for the dynamics. They should therefore be included by some
means if one is trying to mimic the physically relevant degrees of
freedom with the clusters.  A possibility for doing this is to take $M
= h_j$ where $j$ is another randomly chosen point other than the seed
on the surface, which was called the $I$ algorithm ($I=$ integer). This
is no longer ergodic on its own as only even changes in heights will
result from the reflections, but combining the $H$ and $I$ algorithms
gives an ergodic algorithm that effectively eliminates critical slowing
down on fixed lattices.  Another  possibility is to take $M = h_{seed}$
with $q<1$ when $h_i = h_{seed}$ , giving a so-called $Q$ algorithm
which could also be combined with the $H$ algorithm to eliminate
critical slowing down. For the sake of simplicity we concentrate here
on combining the $H$ and $I$ algorithms in our simulations.

In the simulations we measured the width-squared, as defined in
equ.(\ref{e06}), the energy \begin{equation} E =  {\beta \over N} <
\sum_{<ij>} (h_i - h_j)^2 >, \label{e09} \end{equation} the specific
heat \begin{equation} C =   N  (<E^2> - <E>^2), \label{e09b}
\end{equation} autocorrelation functions for the energy and
width-squared and the correlator \begin{equation} C_{ij} = < { 1 \over
n (r) } \sum_{ij}(h_i - h_j)^2  \delta ( d_{ij} - r)>, \label{e10}
\end{equation} where $d_{ij}$ is the geodesic distance on the lattice
between points $i$ and $j$ and \begin{equation} n(r) = \sum_{ij} \delta
( d_{ij} - r) \label{e10a} \end{equation} is the number of points at
geodesic distance $r$.  The size, diameter and number of boundary
points of the two types of clusters and various characteristics of the
lattice itself were also measured. In what follows we describe the
results of these measurements.

\section{Measurements on the SOS model} The energy and specific heat
are plotted in Fig.1 for the various $\beta$ values and system sizes
simulated.  The curve for the specific heat looks gratifyingly KT-like
- there is no sign of an increasing peak with system size as one would
expect for a second order transition. It is also clear that the
behaviour of the model as $\beta \rightarrow 0$ is correctly being
reproduced. In this limit the diverging width of the surface relative
to the lattice size means that one has in effect real valued heights at
each point instead of integers. We should thus see the behaviour of a
single free real scalar field as $\beta \rightarrow 0$ and it is known
from a simple scaling argument that both the energy and specific heat
for this should be $1/2$. Both the specific heat and energy curves in
Fig.1 show this behaviour up to $\beta \simeq 1.5$.  It is interesting
to note that the behaviour of the specific heat curve is actually
better (ie more KT-like) in the SOS model than in the $XY$ model
simulations that we carried out earlier, again with a Wolff cluster
algorithm and similar statistics. Direct measurements in this case
suggested a specific heat curve that was increasing with $N$, whereas
numerical differentiation of the energy curve did not \cite{8}. The
phase transition point is not at the maximum in the specific heat curve
for a KT transition but using this as a rough guide, along with the
onset of the divergence in the width-squared the critical region is in
the region of, or just below, $\beta=3$.

The scaling relation of equ.(\ref{e07}) for the width-squared, which is
plotted in Fig.2 with the smallest $\beta$ values dropped to avoid an
overly large scale is also verified. In Table 1 we show the
results of fitting $\sigma^2$ to $A \log (N) + B$. As might be expected
the fits are quite good deep in the rough phase but deteriorate as one
approaches the transition point in the region of $\beta=3$. We do not
attempt to fit $A ( = {T_{eff} / d \pi}) $ to $T_{eff} - 2 / \pi \simeq
( \beta_c - \beta )^{1/2}$ because of the aforementioned necessity of
including corrections to this formula by introducing a ``running
temperature'' in the $KT$ flow equations. We do not have sufficient
data near to the putative critical point from the current runs to fit
accurately to this modified formula for the width-squared.  {
\begin{center}
\begin{tabular}{|c|c|c|c|c|c|c|c|c|}
\hline $\beta$ &
0.05  & 0.10 & 0.25 & 0.50 & 0.75 & 1.00 & 1.25 & 1.50   \\[.05in]
\hline A        & 11.8(1)   & 6.57(8)    &  2.36(4) & 1.16(2) & 0.83(1)
& 0.61(1)  & 0.50(1) & 0.36(1)\\[.05in] \hline B        & -35(1)    &
-22(1)     &  -7.0(3) & -3.4(1) & -2.6(1) & -1.9(1)  & -1.60(4) &
-0.95(2)\\[.05in] \hline $\chi^2$   & 7.3       & 3.9        &  2.6
& 2.5    & 4.1     & 5.9      & 4.4      & 5.1     \\[.05in] \hline
\end{tabular} \end{center}
%\vspace{.1in}
\begin{center}
\begin{tabular}{|c|c|c|c|c|c|c|} \hline $\beta$ & 1.75  & 2.00 & 2.25 &
2.50 & 2.75 & 3.00    \\[.05in] \hline A        & 0.315(3)  &
0.270(3)   &  0.256(3) & 0.207(3)& 0.153(1)& 0.091(1) \\[.05in] \hline
B        & -0.89(2)  & -0.76(2)   &  -0.85(2)& -0.67(2)& -0.47(1) &
-0.23(1) \\[.05in] \hline $\chi^2$   & 4.3       & 4.3        &
6.5     & 5.7    & 7.3    & 5.9     \\[.05in] \hline \end{tabular}
\end{center}
\vspace{.1in}
\centerline{Table 1: Fitted values of $A$ and $B$.}
\bigskip }

As well as measuring the behaviour of the DGSOS model itself it is also
interesting to investigate the algorithmic aspects of the simulation,
in particular whether the mixed $H$ and $I$ cluster methods employed
here are still efficient at reducing critical slowing down. In the case
of fixed lattices it was found \cite{10} that combining the $H$ and $I$
algorithms gave a very small, or perhaps even zero, value of the
dynamical critical exponent $z$ defined by $\tau \simeq L^z$, where $L$
is the linear size of the system and $\tau$ is the autocorrelation
time, for $\beta \le \beta_c$ \footnote{We restrict ourselves to the
rough phase as the correlation length $\xi$ is effectively the linear
lattice size here - in the smooth phase we would have to fit to the
exponential decay of a two-point function to obtain $\xi$.}. In our
case we must use the definition $\tau \simeq N^{z/d}$ where $d$ is
again the fractal dimension of the lattice We find that $z/d$ is very
small for all of the $\beta$ values we simulated, some typical values
are listed in Table 2.  \begin{center} \begin{tabular}{|c|c|c|c|c|c|}
\hline $\beta$ & 0.05  & 0.25 & 2.50  & 2.75 & 3.00    \\[.05in] \hline
$z/d$        & 0.0018(1)  & 0.0007(1)   & 0.0018(1)  & 0.0089(3) &
0.0027(1) \\[.05in] \hline $\chi^2$   & 2.9       & 4.3        &
10     & 9    & 9      \\[.05in] \hline \end{tabular} \end{center}
\centerline{Table 2: Values of $z/d$ for selected $\beta$.} \bigskip If
we assume that the fractal dimension is $\simeq 2.6$, which is what we
measure in a naive counting of the density of points (see section 4
below), we find values for $z$ that are very small for all the $\beta$
we have measured. We have not, of course, pinpointed the transition
point accurately with the current batch of simulations so it would be
unwise to claim that critical slowing down is almost eliminated on
dynamical lattices with the cluster algorithms used here - we would
need data at the critical point to assure us of this. Nonetheless, it
would appear that a small value of $z$ is likely to be achieved there
too, judging from the persistently low values for all the measured
$\beta$.

The behaviour of the cluster algorithms as $\beta$ is varied is shown
in Fig.3, where the average cluster size as a fraction of the total
number of lattice points is plotted.  In the critical region ($\beta
\simeq 3$) both the $H$ and $I$ algorithms generate clusters of
approximately the same {\it average} fractional size. This is slightly
misleading as a closer look at the cluster statistics reveals that the
$I$ algorithm produces more clusters of intermediate size than the $H$
algorithm, which tends to favour both very large and very small
clusters. This behaviour is similar to that of the clusters on a fixed
lattice.

There is an interesting crossover in the average cluster sizes of the
two algorithms in the region of the phase transition point.  Deep in
the rough phase at low $\beta$ the $H$ algorithm produces a small
average cluster size with fewer clusters generated per attempt.  This
is because at low $\beta$ there is a good chance that the random
reflection plane picked will be a long way from the seed point in the
$H$ algorithm, so the cluster growing will fail at the first step
because of the large penalty in the probability factor. The $I$
algorithm, on the other hand, always picks a point at distance $1/2$
from the seed and incurs no such penalty. In the smooth phase ($\beta
\ge 3$) the situation is reversed - we now see that the $H$ algorithm
generates clusters almost the size of the graph, whereas the $I$
algorithm has a small average size. The histogram of the cluster sizes
reveals that the $H$ algorithm produces clusters almost the size of the
graph at most attempts, whereas the $I$ algorithm has a very much
smaller peak for clusters of this size and fails to build a cluster
with much greater probability. This is presumably because the $I$
algorithm is more likely to encounter a point on the opposite side of
the reflection plane for a thin surface at an early stage and terminate
the cluster growth.

We have also measured the size and diameter of the clusters grown and
binned values for these on an $N=10000$ lattice, without error bars for
clarity, are plotted against each other on log-log scales for three
$\beta$ values in Fig.4. The $H$ and $I$ clusters behave differently
with temperature - the curves for the $I$ clusters shift monotonely
upwards with temperature, but the highest curve for the $H$ clusters is
roughly in the critical region at $\beta=2.75$. In spite of this
variation the slopes in the linear section of the graphs are almost
constant, which gives a fractal dimension (with a large error due to
the spread in the individual clusters) of 2.5 similar to that measured
naively for the graphs themselves as discussed in section 4 below.

The results of the measurements on the correlator $C_{ij}$ are plotted
in Fig.5 for $\beta=0.05$. As we are deep in the rough phase we would
expect essentially free-field behaviour, which naively suggests a
logarithmic growth with $d_{ij} = |i - j|$. However, we see that the
growth is actually {\it linear} after an initial logarithmic section
which increases with the lattice size.  The form of the curve is
similar on fixed lattices where the flips are switched off, even though
the slope changes slightly, so we are clearly looking at a finite size
effect caused by the spherical topology. Further evidence for this
interpretation is provided by simulating a single free scalar field
with a gaussian action, which gives identical behaviour to the $h$, as
it should for small $\beta$.  Although the complicated fractal nature
of the $\phi^3$ graphs makes naive geometrical arguments rather
dangerous, it is possible to construct a plausibility argument for such
behaviour:  the scalar propagator (ie effectively our correlator) on a
sphere can be written as $\log |z|$ where $|z|$ is the
stereographically projected distance from the sphere on the complex
plane. This is related to the distance $d_{ij}$ we measure on the
sphere by \begin{equation} { | z | \over 2 R}  = { ( d_{ij} / \pi R )
\over \sqrt{ 1 - (d_{ij} / \pi R )^2 } } \end{equation} where $R$ is
the radius of the sphere. If $\log | z (d_{ij}) |$ is plotted as a
function of $d_{ij}$, which is what we are doing when we measure the
correlator, it has a long quasi-linear portion after the initial
logarithmic increase.

\section{Lattice Properties}

The results of the measurements of the lattice properties are
succinctly summarized in Fig.6, which should be compared with the
similar Fig.5 in our earlier simulations of the $XY$ model \cite{8}.
The measured quantities $AL$ and $AF$ relate to the acceptance of the
flip moves on the graphs. A flip can be forbidden either from
constraints arising from the graph (ie no tadpoles and no self-energy
bubbles) or from the energy change in the spin model induced by the
reconnection of the vertices. $AL$ measures the fraction of randomly
selected links which pass the first test and could be flipped according
to the graph constraints and $AF$ measures the fraction of the links
satisfying the graph constraints that actually are flipped, ie pass the
Metropolis test using the DGSOS model energy change. Also plotted is
the fraction of rings of length 3, $PR3$, which serves as an indicator
of the local curvature distribution in the $\phi^3$ graph.

{}From Fig.6 it is clear that $PR3$ has a modest peak in the region of
the phase transition, with a
dip appearing in $AL$ at the same point. This behaviour appears to be
generic in all the models on dynamical lattices that we have simulated.
However, there is no sign of the universality with the central charge,
$c$, of the various lattice properties that we found in the Potts model
simulations \cite{4a} where the curves of $AL$, $AF$ and $PR3$ as
functions of the reduced temperature depended only on $c$ and the
maximum and phase transition values of $PR3$ grew linearly with $c$.
This is not surprising as the $XY$ model also failed to show these
properties.  It is interesting to note, however, that the DGSOS model
displays clearly different lattice properties to even the $XY$ model.
In particular $AF$ is  monotone increasing for the DGSOS model whereas
it is rather similar in form to $AL$ in the $XY$ model with a dip
occurring well below the phase transition point. In view of the duality
between the $XY$ and an (admittedly slightly different) SOS model one
might have expected simply reversing left and right in Fig.6 to have
reproduced the $XY$ model results. In addition the maximum value of
$PR3$ observed for the DGSOS model is 0.2211 at $\beta= 2.00$, which
should be compared with 0.2185 for the $XY$ model on graphs of a
similar size.

We have also attempted a naive measurement of the fractal dimension of
the lattice by simply counting the number of points $N(r)$ within a
given geodesic distance $r$ of a random starting point $i$
\begin{equation} N(r) = < \sum_{ij} \theta ( r - d_{ij}) > \label{e99}
\end{equation} and using $\log N(r) \simeq d \log r$ to extract the
fractal dimension $d$.  A similar procedure gave values in the region
of 2.7 for Potts models when extrapolated to infinite lattice size and
2.6 for the $XY$ model when extrapolated in a self-consistent fashion
\cite{4a,8}. Applying the same methods to the DGSOS model lattices
again produces a value in the region of 2.6. However, the recent work
of Kawai et.al. on a transfer matrix formalism for pure 2d gravity
\cite{kw} shows that a direct measurement of the density of points such
as we have carried out here contains non-universal, lattice dependent
factors \footnote{This accounts for the earlier results of \cite{am},
which failed to find {\it any} fractal dimension on very large bare
lattices.}.  This is likely to remain true with the introduction of
matter so, while our results may be useful for comparing the behaviour
of different models on similar lattices, it is unclear how reliable
they are as measurements of the ``dimension'' of the lattices.

\section{Conclusions}

We have simulated a DGSOS model on dynamical $\phi^3$ graphs using a
mixed cluster algorithm.  There is good evidence for the logarithmic
growth of the width-squared of the surface in the rough phase predicted
by $KT$ theory. The energy and specific heat of the DGSOS model also
have the expected forms and low $\beta$ limits. Measurements of the
dynamical critical exponent suggest that critical slowing down may well
be eliminated completely, though it would take an accurate
determination of the critical point to be absolutely certain of this.
The lattice properties we observed are {\it not} a simple inversion of
those seen in the $XY$ model, which is dual to a slightly different
sort of SOS model. It would be interesting to obtain more data in the
critical region with the DGSOS model and fit to the improved formula
for the width-squared in order to determine the critical $\beta$
accurately. It would also be interesting to look at the dual to the
$XY$ model to see if the lattice properties there are then closer to a
straightforward reflection of those in the $XY$ model. As SOS surfaces
are in effect gauss parametrizations of 2d surfaces embedded in 3d with
no overhangs it would be very useful if cluster algorithms similar to
those used here could be devised for the more complicated actions
including extrinsic curvature that are used in simulations of ``real''
surfaces and string theory.

\section{Acknowledgements} This work was supported in part by NATO
collaborative research grant CRG910091 (CFB and DAJ) and by ARC grant
313-ARC-VI-92/37/scu (WJ and DAJ).  CFB is supported by DOE under
contract DE-FG02-91ER40672 and by NSF Grand Challenge Applications
Group Grant ASC-9217394, and WJ
thanks the Deutsche Forschungsgemeinschaft for a Heisenberg fellowship.
DAJ is supported at LPTHE
by an EEC HCM fellowship. The simulations were carried out on
workstations at Heriot-Watt University and Mainz.

\vfill \eject  \vfill \eject
\centerline{\bf Figure Captions} \begin{description}

\item[Fig. 1.] The specific heat and energy for the various graph sizes
plotted against $\beta$.  \item[Fig. 2.] The width-squared of the
surfaces as defined in equ.(9). The lowest $\beta$ values are not
plotted in order to avoid compressing the scale.  \item[Fig. 3.] The
mean cluster sizes as a fraction of the lattice size for the two
cluster algorithms on the largest ($N=10000$) graphs simulated.
\item[Fig. 4.] Binned sizes of both $H$ and $I$ clusters on $N=10000$
lattices plotted against their diameters on a log/log scale for
selected $\beta$.  \item[Fig. 5.] The correlator $C_{ij}$ at
$\beta=0.05$ for various lattice sizes.  Additional simulations of a
{\it real} gaussian scalar field $X$ on a dynamical lattice with 1000
points and an $h$ field on a {\it fixed} lattice with 500 points are shown as
$N1000X$ and $N500F$ respectively.  \item[Fig. 6.] AF, AL and PR3 for
$N=10000$ simulation; the y-scale applies to AF only, AL and PR3 have
been scaled appropriately to fit on plot.  \end{description}
\end{document}